# Analysis of Rainfall records in India: Self Organized Criticality and Scaling

A. Sarkar and P. Barat

Variable Energy Cyclotron Centre

1/AF Bidhan Nagar, Kolkata 700 064, India

*Abstract:*

The time series data of the monthly rainfall records (for the time period 1871-2002) in All India and different regions of India are analyzed. It is found that the distributions of the rainfall intensity exhibit perfect power law behavior. The scaling analysis revealed two distinct scaling regions in the rainfall time series.

# I. INTRODUCTION

Rainfall is an end product of a number of complex atmospheric processes, which vary both in space and time [1]. It can be considered as one of the dominant factor of the meteo-climatic features of an investigated area. The study of rainfall within a given catchment region can be utilized for several purposes, including hydrological structure design, flood prevention etc. Rainfall has been long analyzed by means of standard statistics such as average value, variance, coefficient of variation and percentiles [2]. Recently, Peters et. al. [3] have presented a power law behavior in the distribution of rainfall over at least four decades. In their article, Peters et. al. [3] speculate that such power law behavior indicates that rainfall arises due to Self Organized Criticality (SOC) [4,5] of the system of the atmosphere.

The concept of SOC was introduced by Bak et al. [4] to explain the behavior of open and extended driven systems with avalanche-like energy dissipation, as sandpile [6] and earthquake [7] model. This concept is deeply related to the idea of scale invariance of the distribution of relaxation events and with the existence of self-regulatory internal mechanisms that drive the system spontaneously to a statistical stationary state. Many actual systems and phenomena including earthquake [8], world wide web [9], distribution of family names [10] are claimed to fitted into the SOC framework. The presence of power law of the observed distribution of magnitude of events represents a necessary condition for the existence of SOC. The knowledge of the dynamics of the process is decisive for the characterization of the possible SOC behavior [11].

The extreme complexity of atmospheric processes results from the coupling of several non-linear processes having completely different temporal and spatial characteristic scales. In the rainfall events, there are condensation effect (with length scales of order $10^{-5}$ m) together with planetary transfer of air masses and moisture (of order $10^6$ m)[12]. Despite the richness of individual events, several properties of the statistical distributions of suitable meteorological fields are claimed to obey power law tails. This is the main reason why several weather phenomena have been analyzed within the SOC framework.

The microscopic theories of drop growth processes and precipitation share several similarities with the most common SOC models, including the presence of water pumping and release phases with an avalanche character. Thus, we concentrate our attention on the analysis of rainfall regimes [14], exploring the properties of the distributions in regions with different climatic conditions. With this choice we can discuss whether the SOC assumption for precipitation events would be relevant for any region of the world and to what extent it is affected by the local differences of climate conditions.

## II. SOC AND RAINFALL EVENTS

The laws of evolution for systems with SOC are usually very simple and mimic two common properties (i) the ability of storing some amount of energy (or mass) for some time and (ii) the sudden energy release if a certain threshold is reached. These features lead to the following dynamical behavior: The system usually organizes itself towards a state which is unstable in its immediate vicinity. When this neighborhood is reached the system releases energy pushing itself away from that locally unstable state. Such avalanches, which may be characterized by the energy they release, close a cycle which is then repeated over and over again. Occurrence of SOC is related to this kind of statistical steady-state behavior. It is characterized by hyperbolic or power law distributions of several quantities: that of the relative frequency N of avalanches releasing energy (or mass) E,

i.e. $N(E) \sim (1/E)^{\alpha}$ (1)

These distribution laws reflect the absence of any characteristic length or time scale for that particular dynamics, which is referred to as the celebrated scale invariance of SOC systems.

In this paper we investigate the evidence for SOC behavior in long-term data sets of rainfall in different regions of India. In particular, we search for hyperbolic or power law distributions of extreme events. Though not entirely conclusive, the presence of such a scaling signature will strengthen the link between rainfall dynamics and SOC. The existence of such a link can be inferred also from theoretical arguments, both of global and microphysical type: The overall balance of water content in the atmosphere is governed by evaporation from subtropical oceans and transport within the atmosphere. After a certain residence time in the atmosphere, this water is released in an avalanche-like event. The specific dynamics of such an event is largely controlled by the physics of drop growth and subsequent precipitation [13]. After nucleation, cloud droplets grow by diffusion of water vapor to it surpassing the evaporation due to air flow around the droplets. Generally, clouds exhibit a spectrum of different-sized droplets which broadens toward larger droplets as a result of turbulent motion leading to collision and coalescence. The different settling rates of the different-sized droplets cause the collisions to occur. Eventually, cloud droplets are large enough for gravitation to overcome friction and updraft of air masses. This picture is quite similar to the one employed in avalanche models quoted above, and it is corroborated by recent water avalanche experiments [14].

Related arguments have been used to explain the formation of droplets around condensation nuclei and to establish the link between SOC and air humidity fluctuations [15].

## III. ANALYSIS OF DATA SET

In order to attain a comprehensive picture, we have analyzed long-term monthly rainfall records of weather stations around the India. We have analyzed the monthly rainfall signals for All India and separately for five regions of India, namely Northwest (NW), Central Northeast (CNE), Northeast (NE), West Central (WC) and Peninsular (PE) for the period 1871-2002. Fig. 1 shows the location of the different regions on the Indian map. The data were obtained from the Indian Institute of Tropical Meteorology [16]. Fig. 2 shows a typical segment of the monthly rainfall records for All India. These data sets were scrutinized with respect to rain intensity. The rainfall event can be brought into the context of SOC theory according to the statistical law quoted in the previous section. The first step is the identification of rain events with fluctuations of the water content in the atmosphere, hence with avalanches. The rain intensity, defined as the total monthly amount $r$ of rain collected by a given station, corresponds to the mass flow sliding downhill in a sand pile experiment [6]. A precise identification of the liquid actually released in a single pouring event requires the determination of its spatial and temporal limits. This can be achieved, for instance, by radar sensing, but these data sets are generally rather small and sporadic. By way of contrast, weather station records offer large-scale, coherent and systematic information more suitable for climate characterization, which lead us to a consistent statistical description of the phenomena. In each case, we determine the number of events as a function of a characteristic parameter monthly rain quantity in mm.

Distributions $n(x)$ of events with property $x$ that behave like $n(x) \sim x^{-\tau}$ are most conveniently analyzed with the help of the integrated (cumulative) distribution

$$\overline{N}(x) = \int_x^M n(x)dx \qquad (2)$$

where $M$ is the maximal event encountered in the data set. By using the integrated description instead of histograms we avoid data fluctuations in the low (high) value regime induced by the choice of logarithmic (linear) bins. If $M \rightarrow \infty$ (and if $\tau > 1$) then $\overline{N}(x) \sim x^{-\tau+1}$. Our rainfall data are generally confined to the ranges $1 < t < 1584$ months and $0mm < r < 7000mm$. Therefore, we cannot replace M by $\infty$ in Eq. (2) and obtain

$$N(x) := \overline{N}(x)/x \sim \frac{1}{x^\tau}\left[1 - \left(\frac{x}{M}\right)^{(\tau-1)}\right] \qquad (3)$$

Thus, the log-log plot of $N(x)$ vs. $x$ definitely departs from a straight line as $x$ approaches $M$.

Fig. 3 shows the cumulative distributions for all the studied rainfall data. It is seen that all rainfall record distributions follow power law behavior. The cumulative distributions are fitted with the Eqn. (3). All the data are found to fit very well with Eqn. (3). A typical fit for the All India rainfall data is shown in Fig. 4. From the fit we have obtained the power law exponent $\tau$. The $\tau$ values are 1.00(±0.01), 1.67(±0.03), 1.10(±0.01), 1.00(±0.01), 1.27(±0.01) and 1.00(±0.02) for All India, NW, CNE, NE, WC and PN respectively.

IV. SCALING ANALYSIS

Scaling as a manifestation of underlying dynamics is familiar throughout the physics. It has been instrumental in helping scientists gain deeper insights into problems ranging across the entire spectrum of science and technology, because scaling laws typically reflect underlying generic features and physical principles that are independent of detailed dynamics or characteristics of particular model. Scale invariance seems to be widespread in natural systems. Numerous examples of scale invariance properties can be found in the literature like earthquakes, humidity, networks etc [17-19].

There are various methods of scaling analysis. In recent years, the detrended fluctuation analysis (DFA) proposed by Peng et. al. [20] has been established as an important tool for the detection of long-range correlations in time series with non-stationarities. It has been successfully applied to such diverse field of interest as DNA [20], heart rate dynamics [21], temperature fluctuations [22] etc.

The DFA procedure consists of the following steps. In the first step, the profile

$$Y(k) = \sum_{i=1}^{k} [x(i) - \langle x \rangle] \qquad (4)$$

is determined from the time series $x(i)$ of length N. $\langle x \rangle$ indicates the mean value of $x(i)$'s. Next, the profile $Y(k)$ is divided into $N_n = [N/n]$ non-overlapping segments of equal length $n$. In the next step, the local trend for each segment is calculated by a least-square fit of the data. The y-coordinate of the fitted line is denoted by $Y_n(k)$. Then the detrended time series for the segment duration $n$ is defined as

$$Y_s(k) = Y(k) - Y_n(k) \qquad (5)$$

The root-mean square fluctuation of the original time series and the detrended time series is calculated by

$$F(n) = \sqrt{\frac{1}{N} \sum_{k=1}^{N} [Y(k) - Y_n(k)]^2} \qquad (6)$$

Repeating this calculation over all segment sizes, a relationship between $F(n)$ and $n$ is obtained. If $F(n)$ behaves as a power law function of $n$, data present scaling:

$$F(n) \propto n^{\beta} \qquad (7)$$

Finally the double logarithmic plot of $F(n)$ versus $n$ is used to calculate the slope, which gives the scaling exponent $\beta$.

If $0 < \beta < 0.5$, the time series is long-range anti-correlated; if $\beta > 0.5$, the time series is long-range correlated. $\beta = 0.5$ corresponds to Gaussian white noise, while $\beta = 1$ indicates the $1/f$ noise, typical of systems in a SOC state.

We have applied the DFA method the Indian rainfall record time series data. The result of the analysis is shown in Fig. 5.

Evidently the striking feature is that there are two scaling regions with a discernible bend when the two slopes in the two regions are distinctly different. This feature is found in all the analyzed rainfall data. Since the behavior is universal it is a feature of the rainfall in India.

To quantify the scaling behavior, we perform a linear fit in Region I for $0.5 < \log(n) < 1.05$ and denote the slope by $\beta_1$, and similarly in Region II for $1.95 < \log(n) < 2.55$ with slope denoted by $\beta_2$. For smaller $n$ values, all slopes ($\beta_1$) are around 1, but at larger $n$ values the slope ($\beta_2$) becomes ~0.2. Thus, the rainfall signal behaves similarly to $1/f$ noise for fine temporal resolutions and shows a different correlation structure for coarser resolution.

A change in scaling exponent in physical systems is often attributed to distinct dynamical processes underlying the generation of the time series. An interesting question is how this finding relates to rainfall meteorology.

## V. CONCLUSIONS

We have investigated long time series of the rainfall records for All India and different regions of India and succeeded in finding evidence for power law distributions of the rainfall quantity. This supports the view that atmospheric dynamics is governed, at least in part, by self-organized criticality. The detrended fluctuation analysis revealed two distinct scaling regions in the rainfall records.

**Figure captions**

Fig.1. The map of different regions of rainfall.

Fig. 2 A typical segment of the monthly rainfall records of All India.

Fig. 3 Cumulative distributions of the rainfall records in All India and different regions of India.

Fig. 4 Log-log plot of monthly rainfall distributions for All India.

Fig.5. Detrended fluctuation analysis of the rainfall data.

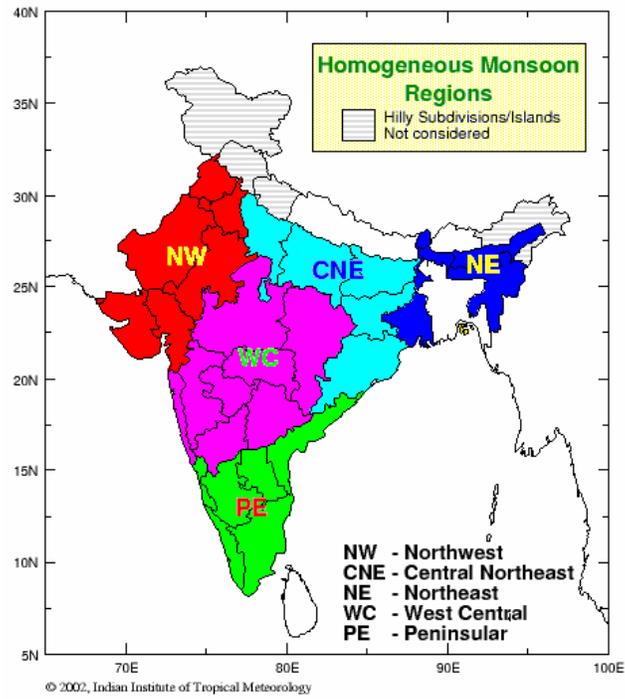

Fig. 1

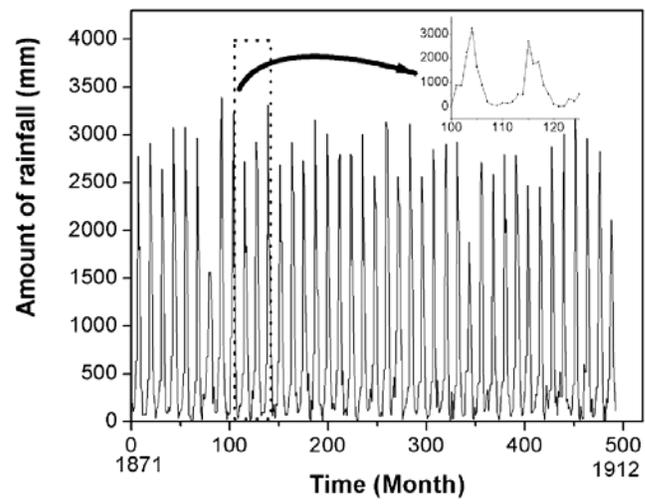

Fig. 2

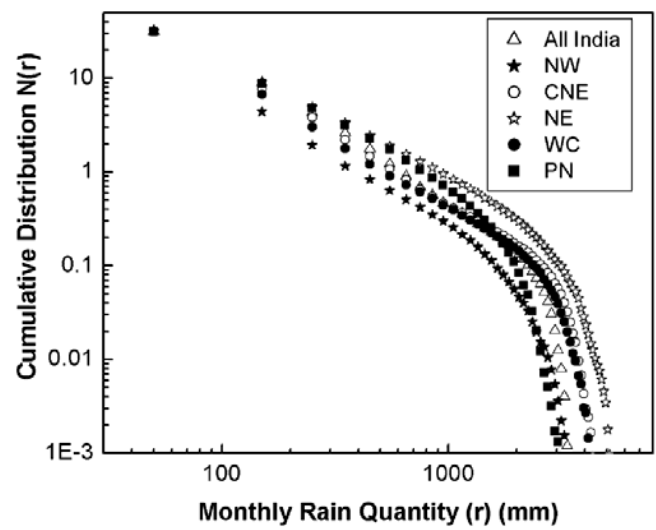

Fig. 3

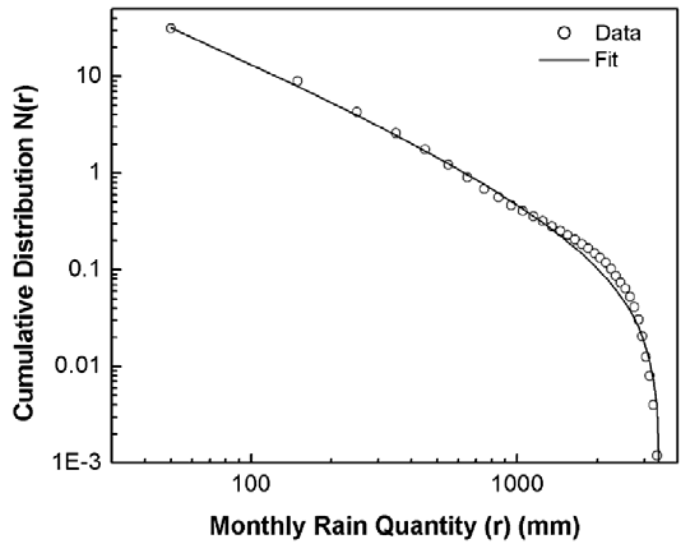

Fig. 4

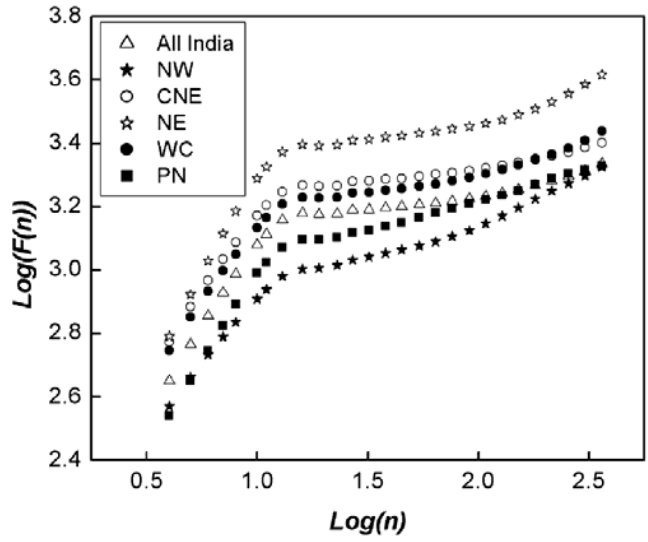

Fig. 5